\documentclass[twocolumn,aps,amsmath,amssymb,prl]{revtex4}
\usepackage{graphicx}
\usepackage{bm}         
\usepackage{color}         
\usepackage{amsmath}
\usepackage{units}
\usepackage{bbm}
\usepackage{textcomp}

\date{\today}
\begin{document}

\title{Dynamical spin polarization of excess quasi-particles {in superconductors} }

\author{Julia~S.~Meyer, Manuel~Houzet}
\affiliation{Univ.~Grenoble Alpes, CEA, IRIG-{Pheliqs}, F-38000 Grenoble, France}
\author{Yuli~V.~Nazarov}
\affiliation{Kavli Institute of NanoScience, Delft University of Technology, Lorentzweg 1, NL-2628 CJ, Delft, The Netherlands}

\begin{abstract}
We show that the annihilation dynamics of excess quasi-particles in superconductors may result in the spontaneous formation of large spin-polarized clusters. This presents a novel scenario for spontaneous spin polarization. We estimate the relevant scales for aluminum, finding the feasibility of clusters with total spin  $S \simeq 10^4 {\hbar}$ that could be spread over {microns}. The fluctuation dynamics of such large spins may be detected by measuring the flux noise in a loop hosting a cluster.
\end{abstract}

\maketitle

Various experiments using superconductors have been interpreted in terms of
a long-lived, non-equilibrium quasi-particle population that persists at low temperatures~\cite{Pekola2000,Aumentado2004, Martinis2009, Catelani2011, deVisser2011, Lenander2011, Peltonen, Pekola, Rajauria2012, Wenner2013, Riste2013, Pekola-review, LevensonFalk2014, Plourde2014, Vool2014, Klapwijk, Wang, Riwar-traps, Pop-preprint, Devoret-preprint, note}. Such quasi-particles may be created, for example,  by Cooper pair breaking due to the absorption of stray photons or cosmic rays -- the dominant mechanism is not clear at the moment. The bottleneck for their evacuation is the two-particle recombination mediated by the electron-phonon interaction. A simple balance predicts a residual quasi-particle density $n\sim c_0= (2A/\bar\Gamma)^{1/2}$, where $A$ is the rate of non-equilibrium generation of quasi-particles per unit volume, and $\bar{\Gamma}$ is a material constant characterizing the inelastic quasi-particle relaxation due to the electron-phonon interaction. The subject has attracted much interest recently as excess quasi-particles will ultimately limit the performance of many superconducting devices~\cite{Martinis2009, deVisser2011, Lenander2011, Rajauria2012, Riste2013, Wenner2013, LevensonFalk2014}. Therefore one needs to deepen earlier studies on quasi-particle relaxation as, e.g., Ref.~\cite{Kaplan1976}. {Several strategies, such as quasi-particle trapping in normal islands or vortices~\cite{Pekola2000,Rajauria2012,Pekola-review,Riwar-traps} and quasi-particle pumping with microwave pulse sequences~\cite{pulses}, can be used to evacuate quasi-particles from the region of interest and lead to a better device performance.} By contrast, unintentional trapping of quasi-particles in bound states below the superconducting gap edge, present in disordered superconductors, may slow down the relaxation dramatically at low concentrations~\cite{bespalov} since the recombination requires two quasi-particles and thus is exponentially suppressed for those in distant bound states.

All above considerations neglect the quasi-particle spin. We note the spin selectivity of the recombination process: in the absence of interactions violating spin conservation, the recombination only proceeds if two quasi-particles are in a spin-singlet state. In this Letter, we show that this spin selectivity may become a mechanism of non-equilibrium spin polarization. The quasi-particles align their spins forming a polarized cluster with greatly enhanced concentration, the number of particles in the cluster and its size being limited by spin relaxation processes. We derive the corresponding conditions for aluminum, showing the feasibility of the clusters of ${\sim} 10^4$ quasi-particles that could be spread over {microns}. The polarization of the cluster slowly fluctuates in time, and we propose a simple setup where the resulting noise can be utilized for the experimental observation of the phenomenon.

{A cluster consists of} an ensemble of quasi-particles with mutually overlapping wavefunctions. In the presence of spin-singlet recombination, a cluster of $N$ quasi-particles is stable only if no pair of quasi-particles has an overlap with a spin-singlet state. This is the case if the cluster is in a maximal spin state, with total spin $S=N/2$. Let us align the $z$-axis with the cluster polarization.

If a new quasi-particle is added to such a cluster, the number of quasi-particles changes by 1: $N\to N'=N+1$, whereas the total spin changes by $\pm1/2$: $S\to S'=S\pm1/2=(N\pm1)/2$. The $z$-projection of the spin is $S_z'=N/2+s_z$, where $s_z=\pm1/2$ is the $z$-projection of the spin of the incoming particle. Thus, if $s_z=1/2$, we obtain the maximal spin state $|S'=(N+1)/2,S_z'=(N+1)/2\rangle$. By contrast, if $s_z=-1/2$, there are two possible spin states: $|S'=(N\pm1)/2,S_z'=(N-1)/2\rangle$. The relative probabilities of these two possibilities are determined by the corresponding Clebsch-Gordan coefficients, which are given in Sec.~I of the Supplemental Material (SM)~\cite{SM}. Note that  $|S'=(N+1)/2,S_z'=(N-1)/2\rangle$ is also a maximal spin state, though {its polarization is not along the $z$-axis anymore.} Since the orientation of the incoming spin is random, the {probabilities for being and not being in a maximal spin state are thus given as {$[1+1/(N+1)]/2$ and $[1-1/(N+1)]/2$}, respectively.} As a consequence, the probability that the new cluster is stable is larger than the probability that the new cluster can decay. This asymmetry thus favors the growth of spontaneously polarized clusters of quasi-particles. The polarization axis of such a cluster is not fixed, but changes randomly and slightly with each new quasi-particle added.

From this consideration, we {c}onstruct a simple model for the spin dynamics of excess quasi-particles. To do so, we consider $N$ quasi-particles in a volume $V$. We {a}ssume that the diffusion of the particles is sufficiently fast that the spatial structure of their wavefunctions does not affect the spin dynamics and concentrate on spin effects only. Let us consider clusters that are close to the stable configuration with maximal spin $S=N/2$. We choose the instantaneous spin quantization axis such that $S_z=S$ and describe the cluster's deviation from the maximal spin state with the integer $m=N/2-S$, $m \ll N,S$. 

{We consider four different processes that can change the state $(N,m)$ of the cluster:}

{(1) Quasi-particle injection:} Quasi-particles are injected with a rate $AV$ and arbitrary spin. Thus, half of them {are} aligned with the polarization axis of the existing cluster{,} whereas half of them are antialigned. If the spin is antialigned, we find that the probability of creating an additional spin flip, $m\to m+1$, is $(N-m)/(N-m+1)$. The possible processes are thus $(N,m)\to(N+1,m)$ with rate $AV[1+1/(N-m+1){]}/2$ and $(N,m)\to(N+1,m+1)$ with rate $AV[1-1/(N-m+1)]/2$.

{(2) Singlet annihilation:} Such annihilation processes are possible only if the system is not in a maximal spin state. At small concentration of spin flips, $m\ll N$, the corresponding rate is, thus, proportional to $m$. In particular, the process $(N,m)\to(N-2,m-1)$ happens with rate $\bar\Gamma (N-m)m/V$.

{(3) Spin flips:} Spin-orbit coupling admits for inelastic spin-flips via the electron-phonon interaction. We assume that each spin may flip independently. As for the injection process, a spin flip does not necessarily change the total spin -- however we will neglect the corresponding $1/N$-corrections to the rates. The rate for the process $(N,m)\to(N,m+1)$ is then given as $(N-m)/\tau_{\rm s}$, where $1/\tau_{\rm s}$ is the spin flip rate for a single spin. Similarly the process $(N,m)\to(N,m-1)$ has the rate $m/\tau_{\rm s}$.

{(4) Triplet annihilation:} In the presence of spin-orbit coupling, pairs of quasi-particles may annihilate even when they are in a spin-triplet state. To account for such processes, we introduce a weak spin-independent annihilation, $\bar\Gamma_{\rm t}\ll\bar\Gamma$. Taking into account all possible orientations of the spins of the annihilated particles, this adds the following processes: $(N,m)\to(N-2,m)$ with rate $\bar\Gamma_{\rm t}(N-m)^2/(2V)$ as well as $(N,m)\to(N-2,m-1)$ with rate $\bar\Gamma_{\rm t}(N-m)m/V$ and $(N,m)\to(N-2,m-2)$ with rate $\bar\Gamma_{\rm t}m^2/(2V)$. 

With this, the dynamics are described by a master equation explicitly given in Sec.~II of the SM~\cite{SM}. As a first step, we derive the mean field solutions for the {most probable} $N$ and $m$. The evolution equations for these quantities read:
\begin{eqnarray}
\frac{dN}{dt}&=&AV-2\frac{\bar\Gamma}V(N-m)m-\frac{\bar\Gamma_{\rm t}}VN^2,\label{eq-N}\\
\frac{dm}{dt}&=&\frac{AV}2\left(1-\frac1{N-m}\right)-\frac{\bar\Gamma}V(N-m)m \label{eq-m}\\
&&-\frac{\bar\Gamma_{\rm t}}VNm+\frac1{\tau_{\rm s}}(N-2m).\nonumber
\end{eqnarray}
Using $\bar\Gamma_{\rm t}\ll\bar\Gamma$ and $m\ll N$, Eq.~\eqref{eq-N} yields the stationary solution
\begin{eqnarray}
m_0&=&\frac{AV^2-\bar\Gamma_{\rm t}N_0}{2\bar\Gamma N_0}.
\end{eqnarray}
Substitution into Eq.~\eqref{eq-m} gives an equation for the average $N$ in the cluster:
\begin{eqnarray}
0&=&\frac{AV}{N_0}-\frac2{\tau_{\rm s}}N_0-\frac{\bar\Gamma_{\rm t}}VN_0^2.\label{eq-mf}
\end{eqnarray}
We can distinguish two regimes, depending on whether spin relaxation (SR) or triplet annihilation (TA) dominates. In the SR regime, Eq.~\eqref{eq-mf} yields $N_0^{\rm (s)}=\sqrt{AV\tau_{\rm s}/2}$, while in the TA regime, one finds $N_0^{\rm (t)}=\left(AV^2/\bar\Gamma_{\rm t}\right)^{1/3}$. The corresponding values for $m$ are given as $m_0^{\rm (s)}=(V/\bar\Gamma\tau_{\rm s})N_0^{\rm (s)}$ and $m_0^{\rm (t)}=[(AV^2\bar\Gamma_{\rm t}^2)^{1/3}/2\bar\Gamma]N_0^{\rm (t)}$, respectively. Comparing the two expressions for $N_0$, we conclude that the TA regime requires $ A>V/(\bar\Gamma_{\rm t}^2\tau_{\rm s}^{3})$, that is, a sufficiently high injection rate at any given volume. 

{The above equations allow us to derive the requirements for the cluster to be highly polarized,} that is, $N_0\gg m_0$. Let us first consider a small $A$ such the cluster is in the SR regime. In this case, a sufficiently small volume $V \ll V_c \equiv \bar\Gamma\tau_{\rm s}$ is required. If at a given $V <V_c$ we increase $A$, and therefore the number of particles in the cluster, we cross-over to the TA regime, and a high polarization persists up to $A \simeq A_c (V_c/V)^2$ with $A_c \equiv \bar\Gamma/(\bar\Gamma_{\rm t}\tau_{\rm s})^2$. This requirement is convenient to express in terms of the number of particles in the cluster, $N \lesssim N_c \equiv \bar\Gamma/\bar\Gamma_{\rm t}$. 

\begin{figure}
\resizebox{.35\textwidth}{!}{\includegraphics{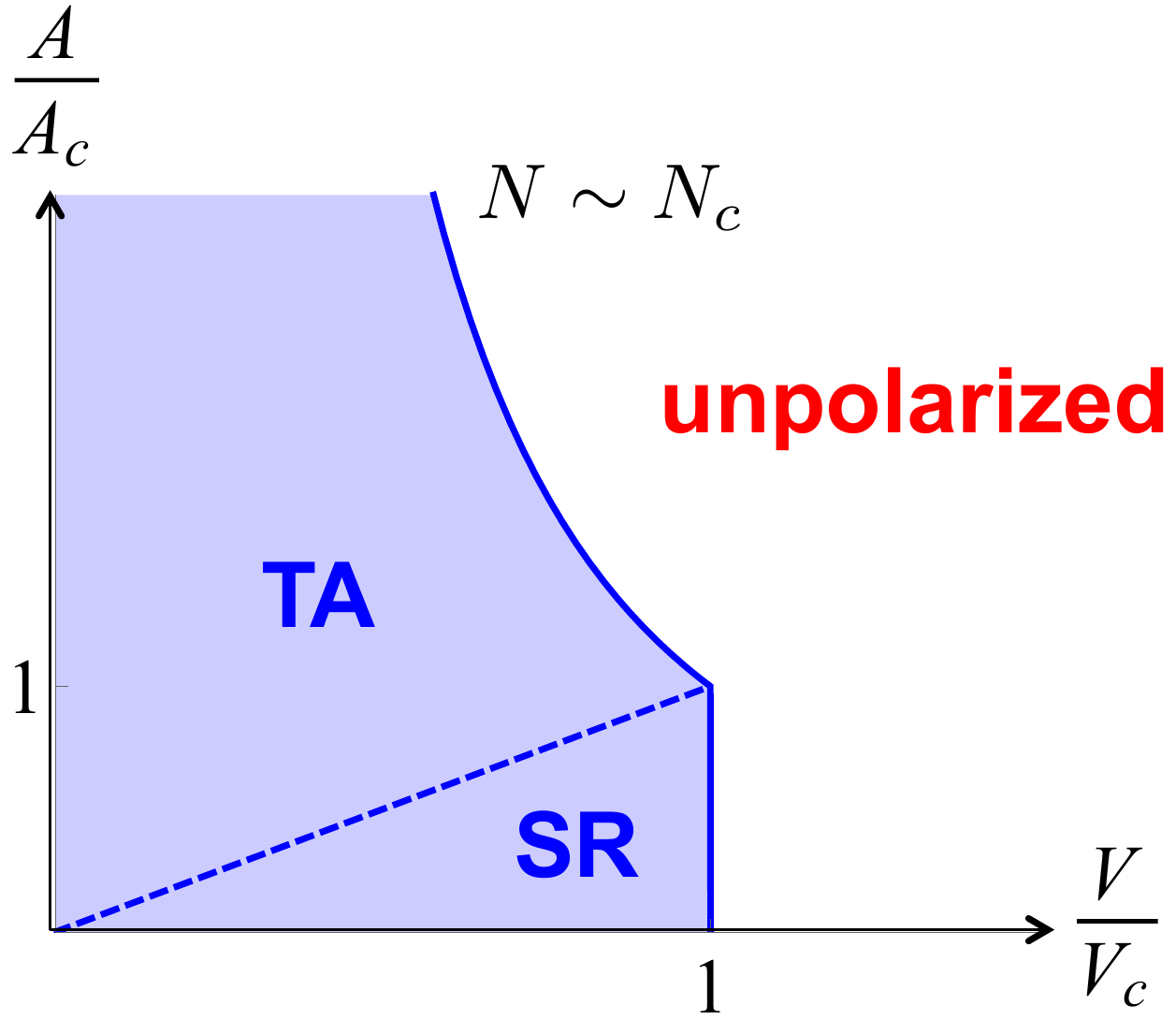}}
\caption{\label{fig-pd} Unpolarized and polarized regimes versus the cluster volume and the injection rate, according to Eqs.~\eqref{eq-res-sf} and \eqref{eq-res-t}. Here $V_c=\bar\Gamma\tau_{\rm s}$,  $A_c=\bar\Gamma/(\bar\Gamma_{\rm t}\tau_{\rm s})^2$, and $N_c=\bar\Gamma/\bar\Gamma_{\rm t}$.}
\end{figure}

Note that, in a polarized cluster, $N_0$ largely exceeds the value $N_{\rm unpol.}=c_0V$ expected for an unpolarized system. It is constructive to express the concentrations as follows: in the SR regime,
\begin{equation}
N_0^{\rm (s)}/V=\frac{c_0}2\zeta_{\rm s}^{-1/2},\quad m_0^{\rm (s)}/V=\frac{c_0}2\zeta_{\rm s}^{1/2},\label{eq-res-sf}
\end{equation}
with $\zeta_{\rm s}\equiv V/V_c {\ll 1}$,
and in the TA regime, 
\begin{equation}
N_0^{\rm (t)}/V=\frac{c_0}2\zeta_{\rm t}^{-1/2},\quad m_0^{\rm (t)}/V=\frac{c_0}2\zeta_{\rm t}^{1/2},\label{eq-res-t}
\end{equation}
with $\zeta_{\rm t}\equiv(AV^2/A_cV_c^2)^{1/3}/2 \ll 1$. The regions where a polarized state is expected are illustrated in Fig.~\ref{fig-pd}, see also Sec.~III of the SM~\cite{SM} for more details.

Let us estimate the relevant material parameters  $\bar{\Gamma}$, $\bar{\Gamma}_{\rm t}$, and $\tau_{\rm s}$. In aluminum, the phonon-assisted recombination rate for quasi-particles near the gap edge is characterized by $\bar\Gamma\simeq 18\, {\rm s}^{-1}\mu {\rm m}^3${~\cite{footnote-Gamma}}. As to the triplet annihilation rate, it involves a phonon emission accompanied by a spin-flip, and is estimated as $\bar{\Gamma}_{\rm t}\sim\alpha_{\rm so}^2 K \bar{\Gamma}$, where $\alpha_{\rm so}\sim 10^{-2}$ is the dimensionless spin-orbit strength, and the suppression factor $K$ reflects the smallness of the momentum transfer in the course of the emission. As such, $K$ crucially depends on the wave vector $q$ of the phonon involved that is set by the energy $\sim \Delta$ released, $cq \simeq \Delta$, $c$ being the sound velocity {and $\Delta$ being the superconducting gap}. In the absence of disorder, $K \simeq (qa)^2$~\cite{Yafet}, $a$ being the interatomic distance scale. With the disorder setting a mean free path $l$, $K \simeq (ql)^{-1}$ for $1 \lesssim ql \lesssim (l/a)^{2/3}$, $K \simeq ql$ for $q \lesssim l^{-1}$ \cite{Schmid}. To have a disorder-independent estimation, we resort to the least suppressed case, $K{=} 1$. This gives $\bar{\Gamma}_{\rm t}\sim\alpha_{\rm so}^2 \bar{\Gamma}\sim 10^{-4} \bar{\Gamma}$. 
 
 It may seem that the relevant spin-flip rate is determined by elastic spin-orbit processes as it is usual in the context of spin transport \cite{spintronics}, $1/\tau_{\rm so} \sim\alpha_{\rm so}^2({\tt \delta\epsilon}/\Delta)^{1/2}/\tau_{\rm el}$, where $\tau_{\rm el}$ is the elastic scattering time, and $\delta\epsilon \lesssim \Delta$ characterizes the energy window for the excess quasi-particles above the superconducting gap~\cite{Zhao1995}. However, this estimation holds for propagating electron waves rather than for the localized states we are dealing with. As explained in \cite{Khaetski}, elastic spin-orbit interaction is inefficient in relaxing the spin of localized states{,} not lifting the Kramers degeneracy. Therefore the spin flips should involve inelastic processes. We assume that the dominant spin-flip process is the phonon emission/absorption in the presence of spin-orbit coupling. The corresponding rate is then estimated as $ 1/\tau_{\rm s}\sim \alpha_{\rm so}^2({\delta\epsilon}/\Delta)^{7/2}K/\tau_0$, where $\tau_0\sim 400\, {\rm ns}$ in Al is the normal-state inelastic phonon scattering time at {energy $\sim\Delta$}~\cite{Kaplan1976}. The first and second suppression factors reflect the smallness of the spin-orbit interaction and the reduction of the density of states \cite{Kaplan1976}, and the factor $K$ now corresponds to the energy transfer $\delta \epsilon \simeq cq$. As above, we resort to the least suppressed choice $K{=} 1$. Even this choice gives very long spin-flip times: at $\delta \epsilon \simeq 0.1 \Delta$ we estimate $\tau_{\rm{s}} \simeq 10 \,\rm{s}$.

With this, we estimate the critical volume $V_c = \tau_{\rm s} \bar{\Gamma}\sim 180\, \mu m^3$. This implies that the spin-polarized cluster can be spread over {micron lengths} and $V_c$ is not a very restrictive parameter. A more severe restriction comes from the triplet annihilation that sets the maximum number of particles in the cluster, $N_c=\bar\Gamma/\bar\Gamma_{\rm t}\sim 10^4$. The critical injection rate, where the cross-over from spin-flip limited to triplet-annihilation limited clusters size takes place, is then estimated as $A_c\sim 10^5\, \rm{s}^{-1}\,\mu {m}^{-3}$. {(A similar injection rate was reported in Ref.~\cite{Wang}.)} The quasi-particle density is enhanced compared to the unpolarized case, if $V<V_c$ and $A<A_c(V_c/V)^2$.

It is important to note that the number of particles in the cluster strongly fluctuates. The mean-field solution gives the most probable number of particles in the cluster, $N_0$, while $\langle N\rangle$ differs from $N_0$ by a factor and the fluctuations $\langle\langle N^2\rangle\rangle=\langle N^2\rangle-\langle N\rangle^2$ are of the order $N_0^2$. To quantify {the fluctuations}, we utilize a Fokker-Planck equation, {cf.~Sec. II of the SM \cite{SM},} that gives the distribution function
\begin{eqnarray}
P(N)=CN^2\exp\left[-\frac{2\bar\Gamma_{\rm t}}{3AV^2}N^3-\frac2{AV\tau_{\rm s}}N^2\right],
\end{eqnarray}
where the constant $C$ ensures the normalization. For the SR and TA regimes, this gives{, respectively,}
\begin{eqnarray}
\label{varianceSF}
\langle N\rangle^{\rm (s)}=\frac2{\sqrt\pi}N_0^{\rm (s)}\!, \quad\langle N^2\rangle^{\rm (s)}=\frac32\left(N_0^{\rm (s)}\right)^2, \\
\label{varianceTA}
\langle N\rangle^{\rm (t)}=\frac{\Gamma\left(\frac 13\right)}{18^{1/3}}N_0^{\rm (t)}\!, \quad \langle N^2\rangle^{\rm (t)}=\frac{2^{4/3}\pi}{3^{5/6}\Gamma\left(\frac 13\right)}\left(N_0^{\rm (t)}\right)^2.
\end{eqnarray}
{A comparison with the classical model~\cite{bespalov} is provided in Sec.~IV of the SM~\cite{SM}.}

\begin{figure}
\resizebox{.35\textwidth}{!}{\includegraphics{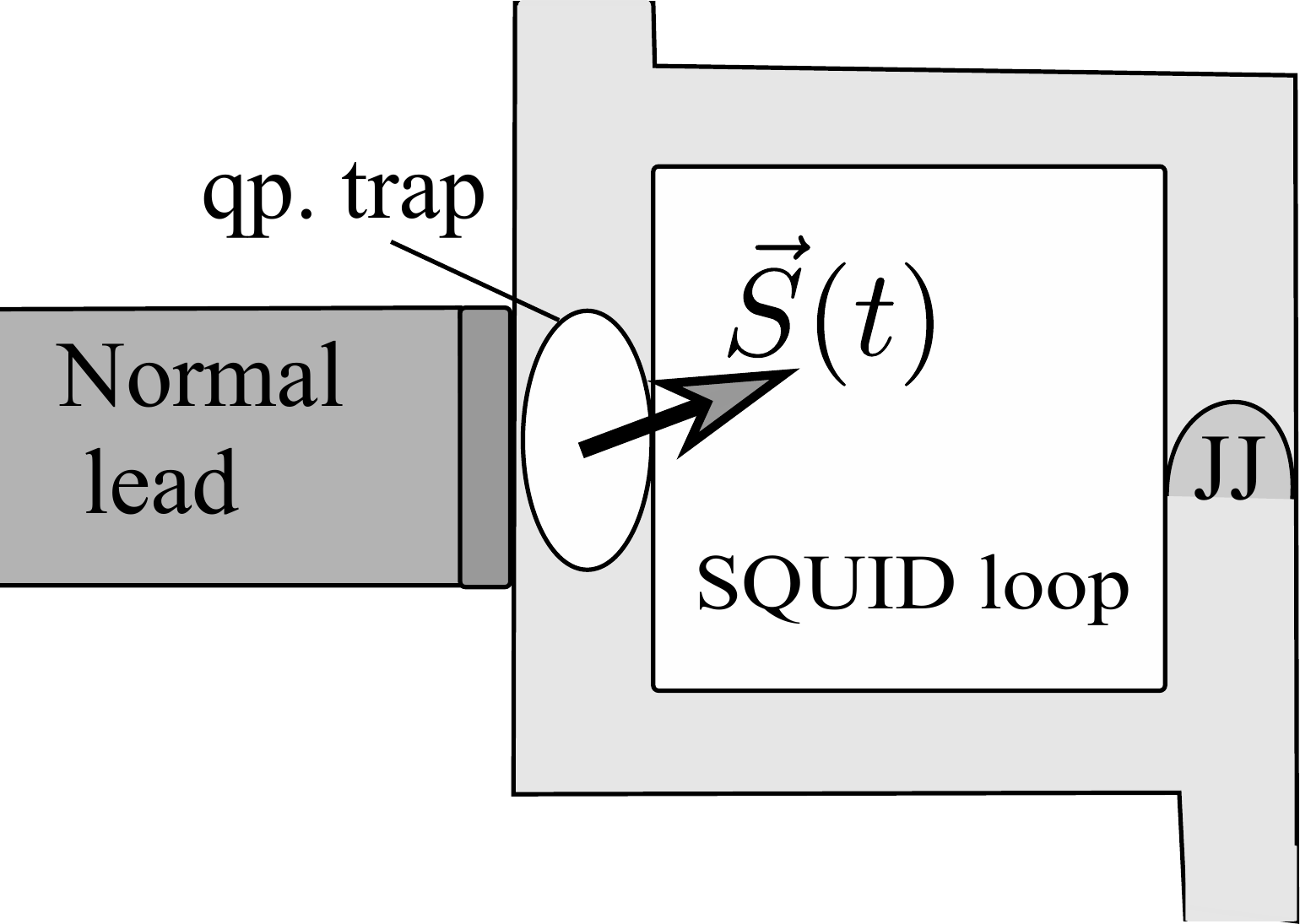}}
\caption{
\label{fig:squid} 
Experimental setup for the {detection} of the polarized state. The quasi-particles are injected from a normal lead into a trap {embedded in the arm} of a SQUID loop. The fluctuations of their common spin produce in the loop the flux noise to be measured. The polarization is seen as an enhanced noise $\propto N_0^2$.}
\end{figure}

While the quasi-particle system is polarized, its spin quantization axis is not fixed but diffuses with time, along with the number of the polarized particles. This produces a measurable spin noise that can be utilized for the experimental verification of the polarization{, using the setup sketched in Fig.~\ref{fig:squid} as explained below}. The spin noise for a certain spin component can be estimated in terms  of the noise of the number of particles $S_N$,  $S_{{\rm spin}} \simeq \tfrac{1}{3} S_{N} \simeq \langle\langle N^2 \rangle\rangle t_f$ . Here, $t_f$ is a characteristic time scale for the fluctuations, that is estimated as {$t_f\simeq N_0 (N_0/AV)$; it yields} $t_f \simeq \tau_{\rm s}$ and $t_f \simeq (A\bar\Gamma_{\rm t}/V)^{-1/3}$ in the SR and TA regimes, respectively. We have evaluated numerically the particle number {zero-frequency} noise in these two regimes to find $S_{N}(0) = 0.5 \langle\langle N^2\rangle\rangle \tau_s$ and $S_{N}(0) = {0.6} \langle\langle N^2\rangle\rangle (A \bar{\Gamma}^2_t /V)^{-1/3}$ where the variances $\langle\langle N^2\rangle\rangle$ in the regimes are given by Eqs. \eqref{varianceSF},~\eqref{varianceTA}.

A flux noise of substantial amplitude $S_{\Phi} \simeq 10^{-12} \,\Phi_0^2 / {\rm Hz}$ at low frequencies is routinely measured in superconducting quantum interference devices (SQUIDs), here $\Phi_0$ is the flux quantum. This noise limits the performance of superconducting qubits, that motivated its thorough investigation~{\cite{Bialczak,Martinis2008, Martinis2012}}. Nowadays its origin is commonly attributed to the slow dynamics of localized spins at the surface of a superconductor~\cite{IoffeNoise, Ioffe2, spin-diffusion}. We note that the spins of non-equilibrium quasi-particles may also contribute to this noise. {In fact, t}he polarization mechanism predicted in this article make these spins very effective noise sources: $N$ quasi-particle spins combined in a polarized cluster produce the same noise as $N^2$ localized spins, provided the time scale of their dynamics is the same. In distinction from localized spins, the quasi-particles can be brought to the superconductor in a controllable way, for instance, by injection through a normal lead separated from the superconductor by a tunnel barrier~\cite{injection}.

This leads us to the suggestion of a concrete experimental setup to observe the {predicted polarized state. As depicted in Fig.~\ref{fig:squid}, o}ne makes a quasi-particle trap {embedded in the arm of a} superconducting loop by reducing the superconducting gap locally, and injects the quasi-particles into the trap from a normal lead that is biased at a voltage that slightly exceeds the reduced gap. The flux noise is monitored at different injection rates corresponding to different quasi-particle numbers $N_0$. {Assuming a width of 100~nm for the SQUID arm, one spin} induces a flux $\simeq 10^{-7} \Phi_0$ through the SQUID loop~{\cite{Bialczak}}. For {the following} estimations, we assume that triplet annihilation dominates, $N_0 = 10^4$, and the trap {volume is $(100\,{\rm nm})^3$}. At these conditions, $t_f \simeq 0.5 \times 10^{-4} \,$s, and the fluctuations of the polarized state produce the noise $S_\Phi \simeq {10^{-12}} \,\Phi_0^2 /$Hz, that exceeds the commonly observed level. If the particles were not polarized, the flux noise would be four orders of magnitude lower. An advantage of the setup is that the number of quasi-particles induced, as well as the fluctuations of this number, can be monitored through the high-frequency inductance and inductance noise of the superconducting sample~\cite{Klapwijk2014}.

In this work, we assume that a possible external magnetic field does not polarize the quasiparticle spins. This is valid provided the corresponding Zeeman energy $E_Z \ll \delta \epsilon$. On the level of the master equation, the polarizing effect of the magnetic field can be taken into account by assigning an anisotropy to the spin relaxation, but we have not investigated this.

In conclusion, we propose a novel scenario for spontaneous spin polarization of a finite system under out-of-equilibrium conditions. We predict that, owing to the spin selectivity of recombination, the excess quasi-particles in a superconductor may spontaneously polarize in clusters. The underlying mechanism differs from that considered in Ref.~\cite{AronovSpivak1980} for homogeneous quasi-particle states. For parameters of Al, such a polarized cluster may contain $10^4$ quasi-particles and spread over {microns}. We show that the polarization can be detected as an excess flux noise.

\acknowledgments
We acknowledge valuable discussions with A. Bespalov in the early stages of this work. This work is supported by the Nanosciences Foundation in Grenoble, in the frame of its Chair of Excellence program, the ANR through the grant ANR-16-CE30-0019 and the European Research Council (ERC) under the European Union's Horizon 2020 research and innovation programme (grant agreement \textnumero 694272).

\end{document}


\widetext

\begin{center}
\textbf{\large Supplemental Materials for \lq\lq Dynamical spin polarization of excess quasi-particles {in superconductors} \rq\rq}
\end{center}

\setcounter{equation}{0}
\setcounter{figure}{0}
\setcounter{table}{0}
\setcounter{page}{1}
\makeatletter
\renewcommand{\theequation}{S\arabic{equation}}
\renewcommand{\thefigure}{S\arabic{figure}}
\renewcommand{\bibnumfmt}[1]{[S#1]}
\renewcommand{\citenumfont}[1]{S#1}

\section{I.~Clebsch-Gordan coefficients}
Upon addition of a quasi-particle to a cluster, the probabilities to realize a state with a total spin $S'$ and spin projection $S'_z$ are determined by the corresponding Clebsch-Gordan coefficients for combining the spin $S$ of the existing cluster with the spin $1/2$ of the incoming quasi-particle [See, e.g., L. D. Landau and E. M. Lifshitz, {\em Quantum Mechanics: Non-Relativistic Theory}, Vol. 3 (3rd ed., Pergamon Press, 1977)]. 
The states we consider, labelled by $N$ and $m$, correspond to $S=S_z=(N-m)/2$. Thus, the relevant Clebsch-Gordan coefficients are the following:
\begin{eqnarray}
\left\langle\frac{N-m}2,\frac{N-m}2;\frac12,\frac12\Big|\frac{N-m+1}2,\frac{N-m+1}2\right\rangle&=&1\\
\left\langle\frac{N-m}2,\frac{N-m}2;\frac12,-\frac12\Big|\frac{N-m+1}2,\frac{N-m-1}2\right\rangle&=&\sqrt{\frac1{N-m+1}}\\
\left\langle\frac{N-m}2,\frac{N-m}2;\frac12,-\frac12\Big|\frac{N-m-1}2,\frac{N-m-1}2\right\rangle&=&\sqrt{\frac{N-m}{N-m+1}}
\end{eqnarray}
The first two lines correspond to processes with $\Delta m=0$ whereas the third line describes a process with $m\to m+1$.

\section{II.~Master equation and Fokker-Planck equation}

The master equation that describes the evolution of the probability $P_{N,m}$, taking into account all processes listed in the main text, is
\begin{eqnarray}
\label{master}
\frac{d}{dt}P_{N,m}&=&\frac{AV}2\left[\left(1+\frac1{N-m}\right)P_{N-1,m}+\left(1-\frac1{N-m}\right)P_{N-1,m-1}-2P_{N,m}\right]\\
&&+\frac{\bar\Gamma}V\left[(N-m+1)(m+1)P_{N+2,m+1}-(N-m)mP_{N,m}\right]\nonumber\\
&&+\frac{\bar\Gamma_{\rm t}}{2V}\left[(m+2)(m+1)P_{N+2,m+2}+2(N-m+1)(m+1)P_{N+2,m+1}+(N-m+2)(N-m+1)P_{N+2,m}\right.\nonumber\\
&&\left.-N(N-1)P_{N,m}\right]
+\frac1{\tau_{\rm sf}}\left[(m+1)P_{N,m+1}+(N-m+1)P_{N,m-1}-NP_{N,m}\right].\nonumber
\end{eqnarray}

Assuming that the number of particles exceeds the average density of the unpolarized system, we may write a simplified {master} equation for the fully polarized system, $m\to0$, {where singlet annihilation is the dominant process}. In that case, injection of a quasi-particle with opposite spin plus singlet annihilation yields a process $N+1\to N$ with rate $AV/2$. Similarly spin flip plus singlet annihilation yields a process $N+2\to N$ with rate $1/\tau_{\rm s}$. In particular,
\begin{eqnarray}
\label{master-P}
\frac{d}{dt}P_{N}&=&\frac{AV}2\left[\left(1+\frac1{N}\right)P_{N-1}+\left(1-\frac1{N+2}\right)P_{N+1}-2P_{N}\right]\\
&&+\frac{\bar\Gamma_{\rm t}}{2V}\left[(N+2)(N+1)P_{N+2}-N(N-1)P_{N}\right]
+\frac1{\tau_{\rm s}}\left[(N+2)P_{N+2}-NP_{N}\right].\nonumber
\end{eqnarray}

In the continuum limit, the probability $P(N)$ obeys the equation
\begin{equation}
\frac{\partial P}{\partial t}=\frac{AV}2\frac\partial{\partial N}\left[N^2\frac\partial{\partial N}\left(\frac P{N^2}\right)\!+\!\left(\frac{2\bar\Gamma_{\rm t}}{AV^2}N^2
\!+\!\frac4{AV\tau_{\rm s}}N\right)P\right]\!.\!\!\!\!\!\end{equation}
The stationary solution is Eq.~(7) of the main text.

\section{III.~The cross-over to an unpolarized state}
\begin{figure}[h!]
a)\resizebox{.25\textwidth}{!}{\includegraphics{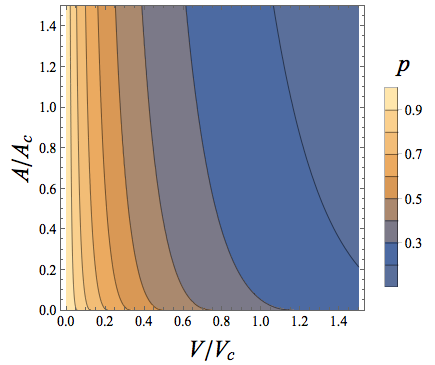}}
b)\resizebox{.26\textwidth}{!}{\includegraphics{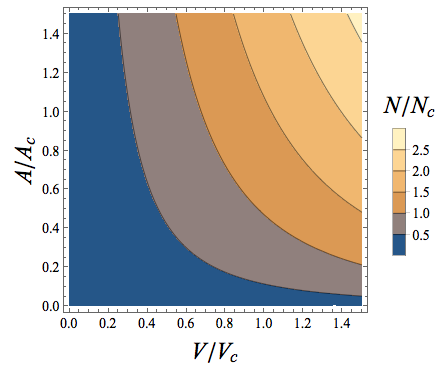}}
c)\resizebox{.28\textwidth}{!}{\includegraphics{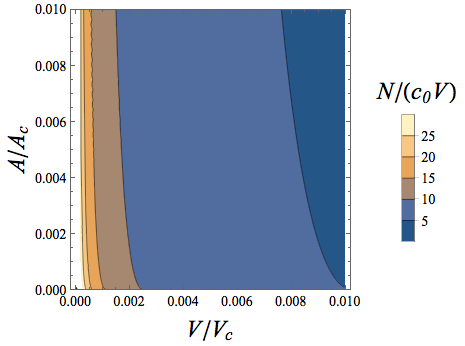}}
\caption{\label{fig-av} {Cross-over between the polarized and unpolarized state as described by Eqs.~\eqref{eq-res-full}.} a) The polarization as a function of $v$ and $a$. b) The number of quasi-particles as a function of $v$ and $a$. c) Detail of the enhancement of the number of quasi-particles compared to the unpolarized case,  $N/(c_0V)$.}
\end{figure}

 While, in principle, our model is only valid at $m\ll N$, we may still use it to better characterize the cross-over to an unpolarized state. The full stationary solution of Eqs.~(3) and (4) of the main text can be written as
\begin{eqnarray}
\frac{AV}{N-m}&=&2\frac {\bar\Gamma}Vm-\frac{\bar\Gamma_{\rm t}}V\frac{N^2}{N-m}\label{stat-2}\\
&=&\left(\frac{\bar\Gamma_{\rm t}}VN+\frac2{\tau_{\rm s}}\right)(N-2m).\label{stat-1}
\end{eqnarray}
If the polarization $p=1-2m/N$ is not too close to 1, namely $1-p\gg N_c^{-1}$, the second term in Eq.~\eqref {stat-2} may be neglected.
Using dimensionless units $a=A/A_c$, $v=V/V_c$, and $n=N/N_c$, we then obtain
\begin{eqnarray}
n=\frac{1-p}p-2v, \qquad a=2(1-p^2)\left(\frac{1-p}{2vp}-1\right)^2.\label{eq-res-full}
\end{eqnarray}
The second equation yields the polarization as a function of $a$ and $v$. Substitution of the result into the first equation then allows one to obtain the number of particles as function of $a$ and $v$. Equations (5) and (6) of the main text are reproduced at $1-p\ll1$. The polarization as well as the number of particles in the $(a,v)$-plane is shown in Fig.~\ref{fig-av}. The polarization smoothly decreases as $a$ or $v$ are increased, see Fig.~\ref{fig-av}a). While $n$ increases with $a$ or $v$, see Fig.~\ref{fig-av}b), the ratio $N/N_{\rm unpol.}$ decreases, see Fig.~\ref{fig-av}c). A large enhancement of the quasi-particle number, $N/N_{\rm unpol.}\gg1$, can be seen at small $a$ and $v$. 

{Within this description, the cross-over between the unpolarized regime and the SR dominated polarized regime  appears smoother than sketched in Fig.~1 of the main text. Namely, according to Eqs.~\eqref{eq-res-full}, a finite polarization $p$ builds up at vanishing injection, $a\to 0$, for a sufficently small volume, $v<v_0$ with $v_0=(1-p)/(2p)$. As the injection increases, the volume corresponding to the same polarization decreases, 
\begin{equation}
a\approx 8p^2\frac{1+p}{1-p}(v_0-v)^2\quad {\rm at}\quad v_0-v\ll v_0.
\end{equation}
}

\section{IV.~Ising spin model}

While the $1/(N-m)$ corrections in the first line of Eq.~\eqref{master} were essential to obtain the mean field result, they are much less important for the fluctuations. To confirm this, we study classical (or Ising) spins as considered in [A. Bespalov {\it et al.}, 
Phys. Rev. Lett. {\bf 117}, 117002 (2016)] for comparison. As the spin quantization axis is now {\lq\lq pinned\rq\rq}, one loses the $1/N$-asymmetry that favors the spontaneous polarization in the quantum (or Heisenberg) case. We note, however, that, if a polarized cluster is generated due to fluctuations, spin relaxation or triplet annihilation are necessary for it to relax, as the singlet recombination process does not change the total spin.

The {m}aster equation in that case is very similar to Eq.~\eqref{master}, however, without the $1/(N-m)$ terms in the first line. Note that, by contrast to the quantum case, where the condition $m\ll N$ was used to estimate the number of available states, here no approximations are necessary: the equation holds for all $m$. Assuming that the number of particles exceeds the average density of the unpolarized system, we may again write a simplified equation for the {polarized system, $m\to0$ (or $m\to N$),} adapting Eq.~\eqref{master-P}{ (i.e., omitting the terms $\sim 1/(N\pm 1)$)}. In the continuum limit, the stationary solution now obeys the equation
\begin{equation}
0=\frac\partial{\partial N}\left[\frac\partial{\partial N}P^{\rm cl}\!+\!\left(\frac{2\bar\Gamma_{\rm t}}{AV^2}N^2
\!+\!\frac4{AV\tau_{\rm s}}N\right)P^{\rm cl}\right]\!,\end{equation}
yielding
\begin{eqnarray}
P^{\rm cl}(N)=C_{\rm cl}\exp\left[-\frac{2\bar\Gamma_{\rm t}}{3AV^2}N^3-\frac2{AV\tau_{\rm s}}N^2\right].
\end{eqnarray}
If spin relaxation dominates, one finds
\begin{eqnarray}
\langle N\rangle_{\rm cl}^{\rm (s)}=\sqrt{\frac{AV\tau_{\rm s}}{2\pi}}.
\end{eqnarray}
If triplet annihilation dominates, one finds
\begin{eqnarray}
\langle N\rangle_{\rm cl}^{\rm (t)}=\frac{6^{1/3}\sqrt\pi}{\Gamma(1/6)}\left(\frac{AV^2}{\bar\Gamma_{\rm t}}\right)^{1/3}.
\end{eqnarray}
Surprisingly the average number of particles $\langle N \rangle_{\rm cl}$ differs from the quantum case only by a prefactor. The same holds true for the fluctuations. By contrast, the {mean field solution} is now given by $N_0^{\rm cl}=N_{\rm unpol.}\ll\langle N \rangle_{\rm cl}$ and $m_0^{\rm cl}=N_{\rm unpol.}/2$.
